\documentclass[aps,onecolumn,showpacs,showkeys,nofootinbib]{revtex4}
\usepackage{epsfig}
\usepackage{amsmath}
\usepackage{amsfonts}
\usepackage{amssymb}
\usepackage{graphicx}
\usepackage{colordvi}
\begin{document}
\title{Comparing scalar-tensor gravity  and $f(R)$-gravity in the Newtonian limit}

\author{S. Capozziello\footnote{e\,-\,mail address:
capozziello@na.infn.it}$^{\diamond}$,\,A. Stabile\footnote{e -
mail address: arturo.stabile@sa.infn.it}$^{\natural}$, \,A.
Troisi\footnote{e\,-\,mail address: antrois@gmail.com}$^{\ddag}$}

\affiliation{$^{\diamond}$ Dipartimento di Scienze Fisiche,
Universit\`a di Napoli "Federico II" and INFN sez. di Napoli
Compl. Univ. di Monte S. Angelo, Edificio G, Via Cinthia, I-80126
- Napoli, Italy}

\affiliation{$^{\natural}$ Dipartimento di Ingegneria, Universita'
del Sannio and INFN sez. di Napoli, gruppo collegato di Salerno,
C-so Garibaldi, I - 80125 Benevento, Italy}

\affiliation{$^{\ddag}$ Dipartimento di Ingegneria Meccanica,
Universita'  di Salerno, Via Ponte Don Melillo, I - 84084 Fisciano
(SA), Italy}

\begin{abstract}
Recently, a  strong debate has been pursued about the Newtonian
limit (i.e. small velocity and weak field) of fourth order gravity
models. According to some authors, the Newtonian limit of
$f(R)$-gravity is equivalent to the one of Brans-Dicke gravity
with $\omega_{BD}\,=\,0$, so that the PPN parameters of these
models turn out to be ill defined. In this paper, we carefully
discuss this point considering that fourth order gravity models
are dynamically equivalent to the O'Hanlon Lagrangian. This is a
special case of scalar-tensor gravity characterized only by
self-interaction potential and that, in the Newtonian limit, this
implies a non-standard behavior that cannot be compared with the
usual PPN limit of General Relativity.
 The result turns out to be
completely different from the one of Brans-Dicke theory and in
particular suggests that it is  misleading to consider the PPN
parameters of this theory with $\omega_{BD}\,=\,0$ in order to
characterize the homologous quantities of $f(R)$-gravity. Finally
the solutions at Newtonian level, obtained in the Jordan frame for
a $f(R)$-gravity, reinterpreted as a scalar-tensor  theory, are
linked to those in the Einstein frame.
\end{abstract}
\date{\today}

\maketitle

\section{Introduction}

Recently,  several authors claimed that higher order theories of
gravity, in particular $f(R)$-gravity \cite{odintsov}, are
characterized by an ill-defined behavior in the Newtonian regime.
In a series of papers \cite{ppn-bad}, it is discussed that higher
order theories violate experimental constraints of General
Relativity (GR) since a direct analogy between  $f(R)$-gravity and
Brans-Dicke gravity \cite{bransdicke} gives the Brans-Dicke
characteristic parameter, in metric formalism, $\omega_{BD}=0$
while it should be $\omega_{BD}\rightarrow\infty$ to recover the
standard GR. Actually despite the calculation of the Newtonian
limit of $f(R)$, directly performed in the Jordan frame, have
showed that this is not the case \cite{noi-prd,altri-ppn-ok}, it
remains to clarify why the analogy with Brans-Dicke gravity seems
to fail its predictions also if one is assuming $f(R)\simeq
R^{1+\epsilon}$ with $\epsilon\rightarrow 0$. The shortcoming
could be  overcome once the correct analogy between $f(R)$-gravity
and the scalar-tensor framework is taken into account.

The  action of the Brans-Dicke gravity, in the Jordan frame,
reads\,:

\begin{equation}\label{bransdicke}\mathcal{A}^{BD}_{JF}=
\int d^4x\sqrt{-g}\biggl[\phi
R+\omega_{BD}\frac{\phi_{;\alpha}\phi^{;\alpha}}{\phi}+\mathcal{X}\mathcal{L}_m\biggr]\,,
\end{equation}
where there is a generalized kinetic term  and no potential is
present.  On the other hand, considering a generic function $f(R)$
of the Ricci scalar $R$, one has\,:

\begin{eqnarray}
\mathcal{A}_{JF}^{f(R)}=\int
d^4x\sqrt{-g}\biggl[f(R)+\mathcal{X}\mathcal{L}_m\biggr]\,.
\end{eqnarray}
In both cases,  $\mathcal{X}=\frac{8\pi G}{c^4}$, is the standard
Newton coupling, $\mathcal{L}_m$ is the perfect fluid  matter
Lagrangian and $g$ is the determinant of the metric.

As said above, $f(R)$-gravity can be re-interpreted as a
scalar-tensor theory by introducing a suitable scalar field $\phi$
which  non-minimally couples with the gravity sector. It is
important to remark that such an analogy holds in a formalism in
which the scalar field displays no kinetic term but is
characterized by means of a self-interaction potential which
determines the  dynamics (\emph{O'Hanlon Lagrangian})
\cite{ohanlon}. This consideration, therefore, implies that the
scalar field Lagrangian, equivalent to the purely geometrical
$f(R)$ one, turns out to be  different with respect to the above
ordinary Brans-Dicke definition (\ref{bransdicke}). This point
represents a crucial aspect of our analysis. In fact, as we  will
see below, such a difference will imply completely different
results in the Newtonian limit of the two models and,
consequently, the impossibility to compare predictions coming from
the PPN approximation of Brans-Dicke models to those coming from
$f(R)$-gravity.

The layout of paper is the following. In Sec. \ref{2}, we discuss the
solutions in the Newtonian limit of $f(R)$-gravity by using the analogies with the O'Hanlon
theory.  Sec.\ref{3} is devoted to the analysis of  the solutions in the limit
$f(R)\rightarrow R$ and the interpretation of PPN parameters
$\gamma$, $\beta$. Conformal transformations and the solutions in the Newtonian limit approximation  are
considered in Sec.\ref{4}. Concluding remarks are drawn in Sec.
\ref{5}.

\section{The Newtonian limit of $f(R)$-gravity by O'Hanlon theory}\label{2}

Before starting with our analysis, let us remind that the  field
equations in metric formalism, coming from $f(R)$-gravity,  are

\begin{eqnarray}\label{fe}
H_{\mu\nu}=f'(R)R_{\mu\nu}-\frac{1}{2}f(R)g_{\mu\nu}-f'(R)_{;\mu\nu}+g_{\mu\nu}\Box
f'=\mathcal{X}T_{\mu\nu}
\end{eqnarray}
which have to be solved together to the trace equation

\begin{eqnarray}\label{trfe}
\Box f'(R)+\frac{f'(R)R-2f(R)}{3}=\frac{\mathcal{X}}{3}T\,.
\end{eqnarray}
Let us notice that this last expression assigns the evolution of
the Ricci scalar as a dynamical quantity. Here, ${\displaystyle
T_{\mu\nu}=\frac{-2}{\sqrt{-g}}\frac{\delta(\sqrt{-g}\mathcal{L}_m)}{\delta
g^{\mu\nu}}}$ is the the energy-momentum tensor of matter, while
$T=T^{\sigma}_{\,\,\,\,\,\sigma}$ is the trace, ${\displaystyle
f'(R)=\frac{df(R)}{dR}}$. The conventions for Ricci's tensor is
$R_{\mu\nu}={R^\sigma}_{\mu\sigma\nu}$ while for the Riemann
tensor is
${R^\alpha}_{\beta\mu\nu}=\Gamma^\alpha_{\beta\nu,\mu}+...$.\\ The
affine connections are the  Christoffel symbols of the metric:
$\Gamma^\mu_{\alpha\beta}=\frac{1}{2}g^{\mu\sigma}(g_{\alpha\sigma,\beta}+g_{\beta\sigma,\alpha}
-g_{\alpha\beta,\sigma})$. The adopted signature is $(+---)$.

On the other hand, the so-called  O'Hanlon Lagrangian
\cite{ohanlon} can be written as

\begin{eqnarray}\label{ohanlon}
\mathcal{A}_{JF}^{OH}=\int d^4x\sqrt{-g}\biggl[\phi
R-V(\phi)+\mathcal{X}\mathcal{L}_m\biggr]\,,
\end{eqnarray}
where $V(\phi)$ is the self-interaction potential. Field equations
are obtained by varying Eq. (\ref{ohanlon}) with respect both
$g_{\mu\nu}$ and $\phi$ which now represent the dynamical
variables. Thus, one obtains

\begin{eqnarray}\label{Eq-bsfR-fe}
\phi
G_{\mu\nu}+\frac{1}{2}V(\phi)g_{\mu\nu}-\phi_{;\mu\nu}+g_{\mu\nu}\Box\phi=\mathcal{X}T_{\mu\nu}\,,
\end{eqnarray}
\begin{eqnarray}\label{Eq-bsfR-fesf}
R-\frac{dV(\phi)}{d\phi}=0\,,
\end{eqnarray}
\begin{eqnarray}\label{Eq-bsfR-KG}
\Box\phi-\frac{1}{3}\biggl[\phi\frac{d V(\phi)}{d
\phi}-2V(\phi)\biggr]=\frac{\mathcal{X}}{3}T\,,
\end{eqnarray}
where we have displayed the field equation for $\phi$.
Eq. (\ref{Eq-bsfR-KG}) is a combination of the trace of
(\ref{Eq-bsfR-fe}) and  (\ref{Eq-bsfR-fesf}).   $f(R)$-gravity
and O'Hanlon gravity  can be mapped one into the other
considering the following equivalences

\begin{eqnarray}\label{tr1}\phi=f'(R)\end{eqnarray}

\begin{eqnarray}\label{tr2}V(\phi)=f'(R)R-f(R)\end{eqnarray}

\begin{eqnarray}\label{tr3}\phi\frac{d V(\phi)}{d\phi}-2V(\phi)=2f(R)-f'(R)R\end{eqnarray}
and supposing that the Jacobian of the transformation $\phi=f'(R)$
is non-vanishing. Henceforth we can consider, instead  of Eqs.
(\ref{fe})-(\ref{trfe}), a new set of field equations determined
by the equivalence between the O'Hanlon gravity and  the
$f(R)$-gravity\,:

\begin{eqnarray}\label{fets}
\phi R_{\mu\nu}-\frac{1}{6}\biggl[V(\phi)+\phi\frac{d V(\phi)}{d
\phi}\biggr]g_{\mu\nu}-\phi_{;\mu\nu}=\mathcal {X}\Sigma_{\mu\nu}
\end{eqnarray}

\begin{eqnarray}\label{fetstr}
\Box\phi-\frac{1}{3}\biggl[\phi\frac{d V(\phi)}{d
\phi}-2V(\phi)\biggr]=\frac{\mathcal{X}}{3}T
\end{eqnarray}
where $\Sigma_{\mu\nu}\doteq T_{\mu\nu}-\frac{1}{3}Tg_{\mu\nu}$.

Let us, now, calculate the Newtonian limit of Eqs.
(\ref{fets})-(\ref{fetstr}). To perform this calculation,  the
metric tensor $g_{\mu\nu}$ and  the scalar field $\phi$ have to
be perturbed with respect to the background.  After, one has to
search for solutions at the $(v/c)^2$ order in term of the metric
 and  the scalar field entries. It is

\begin{eqnarray}g_{\mu\nu}\simeq \begin{pmatrix}
  1+g^{(2)}_{00} & \vec{0}^T \\
  & \\
  \vec{0} & -\delta_{ij}+g^{(2)}_{ij}
\end{pmatrix}\end{eqnarray}
\begin{eqnarray}\phi\sim\phi^{(0)}+\phi^{(2)}\,.\end{eqnarray}
The differential operators turn out to be approximated as

\begin{eqnarray}
\Box\approx\partial^2_0-\Delta\,\,\,\,\,\,\,\,\text{and}\,\,\,\,\,\,\,\,\,_{;\mu\nu}
\approx\partial^2_{\mu\nu}\,,
\end{eqnarray}
since time derivatives increase the degree of perturbation, they can be discarded
\cite{noi-prd}.  From a physical point of view, this position holds since Newtonian limit implies also the slow motion.

Actually in order to simplify calculations, we can
exploit the gauge freedom that is intrinsic in the metric definition. In
particular, we can choose the harmonic gauge
$g^{\rho\sigma}\Gamma^\mu_{\rho\sigma}=0$ so that the components
of Ricci tensor reduces to

\begin{equation}
\left\{\begin{array}{ll}R^{(2)}_{00}=\frac{1}{2}\triangle
g^{(2)}_{00}\\\\R^{(3)}_{0i}=0\\\\R^{(2)}_{ij}=\frac{1}{2}\triangle
g^{(2)}_{ij}\end{array}\right.\,.
\end{equation}
Accordingly, we  develop
 the self-interaction potential at second order. In
particular, the quantities in Eqs. (\ref{fets}) and (\ref{fetstr})
read\,:

\begin{eqnarray}
V(\phi)+\phi\frac{d V(\phi)}{d \phi}\simeq
V(\phi^{(0)})+\phi^{(0)}\frac{d V(\phi^{(0)})}{d
\phi}+\biggr[\phi^{(0)} \frac{d^2V(\phi^{(0)})}{d\phi^2}+2\frac{d
V(\phi^{(0)})}{d\phi}\biggl]\phi^{(2)}\,,
\end{eqnarray}
\begin{eqnarray}
\phi\frac{d V(\phi)}{d \phi}-2V(\phi)\simeq\phi^{(0)}\frac{d
V(\phi^{(0)})}{d \phi}-2V(\phi^{(0)})+\biggr[\phi^{(0)}
\frac{d^2V(\phi^{(0)})}{d\phi^2}-\frac{d
V(\phi^{(0)})}{d\phi}\biggl]\phi^{(2)}\,.
\end{eqnarray}
Field Eqs. (\ref{fets}) - (\ref{fetstr}), solved at 0-th order
of approximation, provide the two solutions

\begin{eqnarray}
V(\phi^{(0)})=0\,\,\,\,\,\,\,\,\text{and}\,\,\,\,\,\,\,\,\,\frac{d
V(\phi^{(0)})}{d \phi}\,=\,0
\end{eqnarray}
which fix the 0-th order terms of the
self-interaction potential; therefore we have

\begin{eqnarray}
-\phi\frac{dV(\phi)}{d\phi}\simeq-\phi^{(0)}
\frac{d^2V(\phi^{(0)})}{d\phi^2}\phi^{(2)}\doteq3m^2\phi^{(2)}\,,
\end{eqnarray}
where the constant factor $m^2$ can be easily interpreted as a mass
term as will become clearer in the following analysis (see also \cite{post-mink}). Now, taking into
account the above simplifications, we can rewrite the field equations
at the  $(v/c)^2$ order in the form\,:

\begin{eqnarray}\label{fets1.2t}
\triangle
g^{(2)}_{00}=\frac{2\mathcal{X}}{\phi^{(0)}}\Sigma_{00}^{(0)}-m^2\frac{\phi^{(2)}}{
\phi^{(0)}}\,,
\end{eqnarray}
\begin{eqnarray}\label{fets1.2r}
\triangle
g^{(2)}_{ij}=\frac{2\mathcal{X}}{\phi^{(0)}}\Sigma_{ij}^{(0)}+m^2\frac{\phi^{(2)}}{
\phi^{(0)}}\delta_{ij}+2\frac{\phi^{(2)}_{,ij}}{\phi^{(0)}}
\end{eqnarray}
\begin{eqnarray}\label{fetstr1.2}
\triangle\phi^{(2)}-m^2\phi^{(2)}=-\frac{\mathcal{X}}{3}T^{(0)}\,.
\end{eqnarray}
The scalar field solution can be easily obtained from Eq.
(\ref{fetstr1.2}) as\,:

\begin{eqnarray}
\phi^{(2)}(\mathbf{x})\,=-\frac{\mathcal{X}}{3}\,\int
d^3\mathbf{x}'\mathcal{G}(\mathbf{x},\mathbf{x}')T^{(0)}(\mathbf{x}')
\end{eqnarray}
where $\mathcal{G}(\mathbf{x},\mathbf{x}')$ is the Green funtion
of the operator $\triangle-m^2$, while, for $g^{(2)}_{00}$ and
$g^{(2)}_{ij}$, we have

\begin{eqnarray}
g^{(2)}_{00}(\mathbf{x})=-\frac{\mathcal{X}}{2\pi\phi^{(0)}}\int
d^3\mathbf{x}'\frac{\Sigma^{(0)}_{00}(\mathbf{x}')}{|\mathbf{x}-\mathbf{x}'
|}+\frac{m^2}{4\pi\phi^{(0)}}\int
d^3\mathbf{x}'\frac{\phi^{(2)}(\mathbf{x}')}{|\mathbf{x}-\mathbf{x}'|}\,,
\end{eqnarray}
\begin{eqnarray}
g^{(2)}_{ij}(\mathbf{x})=&-&\frac{\mathcal{X}}{2\pi\phi^{(0)}}\int
d^3\mathbf{x}'\frac{\Sigma^{(0)}_{ij}(\mathbf{x}')}{|\mathbf{x}-\mathbf
{x}'|}-\frac{m^2\delta_{ij}}{4\pi\phi^{(0)}}\int
d^3\mathbf{x}'\frac{\phi^{(2)}(\mathbf{x}')}{|\mathbf{x}-\mathbf{x}'|}\nonumber
\\\nonumber\\&+&\frac{2}{\phi^{(0)}}\biggl[\frac{x_ix_j}
{|\mathbf{x}|^2}\phi^{(2)}(\mathbf{x})+\biggl(\delta_{ij}
-\frac{3x_ix_j}{\mathbf{|x|}^2}\biggr)\frac{1}{|\mathbf{x}|^3}\int_0^
{|\mathbf{x}|}d|\mathbf{x}'||\mathbf{x}|'^2\phi^{(2)}
(\mathbf{x}')\biggr]\,.
\end{eqnarray}
The above three solutions  are a completely general result
\cite{weinberg}. An  example can make clearer the discussion. We
can consider a fourth order gravity Lagrangian\footnote{It is
important to stress that, in the Newtonian limit of any analytic
$f(R)$-gravity model,  we need to consider only the first two
derivatives of $f(R)$ \cite{noi-prd}.} of the form $f(R)=aR+bR^2$
so that the ``dummy" scalar field reads $\phi=a+2bR$. The relation
between $\phi$ and $R$ is ${\displaystyle R=\frac{\phi-a}{2b}}$
while the self-interaction potential turns out the be
${\displaystyle V(\phi)=\frac{(\phi-a)^2}{4b}}$ satisfying the
conditions $V(a)=0$ and $V'(a)=0$. In relation to the definition
of the scalar field, we can opportunely identify $a$ with a
constant value $\phi^{(0)}\,=\,a$. Furthermore, the scalar field
mass can be expressed in term of the Lagrangian parameters as
${\displaystyle m^2=-\frac{1}{3}\phi^{(0)}
\frac{d^2V(\phi^{(0)})}{d\phi^2}=-\frac{a}{6b}}$. Since the Ricci
scalar at lowest order (Newtonian limit) reads

\begin{eqnarray}
R\simeq
R^{(2)}(\mathbf{x})=\frac{\phi^{(2)}(\mathbf{x})}{2b}=-\frac{\mathcal{X}}{6b}\,\int
d^3\mathbf{x}'\mathcal{G}(\mathbf{x},\mathbf{x}')T^{(0)}(\mathbf{x}')\,,
\end{eqnarray}
if we consider a point-like mass $M$, the energy-momentum tensor
components become respectively $T_{00}=\rho$, $T\sim\rho$ while
$\rho=M\delta(\mathbf{x})$, therefore we obtain

\begin{eqnarray}
R^{(2)}=-\frac{(2\pi)^{1/2}r_gm^2}{a}\frac{e^
{-m|\mathbf{x}|}}{|\mathbf{x}|}\,,
\end{eqnarray}
where $r_g$ is the Schwarzschild radius. The immediate consequence
is that the solution for the scalar field $\phi$, up to the second
order of perturbation, is given by

\begin{eqnarray}
\phi=a+\frac{(2\pi)^{1/2}r_g}{3}\frac{e^{-m|\mathbf{x}|}}{|\mathbf{x}|}\,.
\end{eqnarray}
 In the same way, one can deduce
the expressions for $g^{(2)}_{00}$ and $g^{(2)}_{ij}$, where
$\Sigma^{(0)}_{00}\,=\,\frac{2}{3}\rho c^2$ and
$\Sigma^{(0)}_{ij}\,=\,\frac{1}{3}\rho
c^2\delta_{ij}\,=\,\frac{1}{2}\Sigma^{(0)}_{00}\delta_{ij}$. As
matter of fact, the metric solutions at the second order of
perturbation are

\begin{eqnarray}\label{gsol1}
g_{00}=1-\frac{4}{3a}\frac{r_g}{|\mathbf{x}|}-\frac{(2\pi)^{1/2}}{3a}\frac{r_ge^{-m
|\mathbf{x}|}}{|\mathbf{x}|}\,,
\end{eqnarray}
\begin{eqnarray}\label{gsol2}
g_{ij}=&-&\biggl\{1+\frac{2}{3a}\frac{r_g}{|\mathbf{x}|}-\frac{(2\pi)^{1/2}r_g}{3a}
\biggl[\biggl(\frac{1}{|\mathbf{x}|}-\frac{2}{m|\textbf{x}|^2}-\frac{2}
{m^2|\textbf{x}|^3}\biggr)e^{-m
|\mathbf{x}|}-\frac{2}{m^2|\textbf{x}|^3}\biggr]\biggr\}\delta_{ij}\nonumber\\\nonumber
\\&+&\frac{2(2\pi)^{1/2}r_g}{3a}\biggl[\biggl(\frac{1}{|\textbf{x}|}+\frac{3}{m|\textbf{x}|^2}
+\frac{3}{m^2|\textbf{x}|^3}\biggr)e^{-m|\textbf{x}|}-\frac{3}{m^2|\textbf{x}|^3}\biggr]\frac{x_ix_j}{|\textbf{x}|^2}\,.
\end{eqnarray}
These quantities show that the gravitational potential coming from the O'Hanlon theory of gravity
 is non-Newtonian. The correctios have the meaning of scale parameters defining characteristic  sizes and masses
 \cite{noi-prd, post-mink}.

\section{The behavior of solutions for $f(R)\rightarrow R$ and the interpretation of PPN parameters $\gamma$, $\beta$}\label{3}

The results (\ref{gsol1}) - (\ref{gsol2}) are  equivalent to those obtained in
$f(R)$-gravity  \cite{noi-prd}. This point is   very important since
 such a behaviour prevents
from obtaining the standard definition of the PPN parameters as
corrections to the Newtonian potential. As matter of fact, at the Newtonian level,   it
is indeed not true that a generic $f(R)$-gravity model corresponds
to a Brans-Dicke model with $\omega_{BD}\,=\,0$.   In particular, in such a limit,  it is not correct to  consider the PPN parameter
$\displaystyle\gamma\,=\,\frac{1+\omega_{BD}}{2+\omega_{BD}}$ (see
\cite{will}) of Brans-Dicke gravity and evaluating this at
$\omega_{BD}\,=\,0$. In this case,   one obtains  $\gamma\,=\,1/2$ as suggested
in \cite{ppn-bad} and the standard Newton potential ($\gamma\,=\,1/2$) could never be recovered.   Differently, because of the presence of the
self-interaction potential $V(\phi)$ in the O'Hanlon theory,
 a Yukawa-like correction appears  and it
contributes in a completely different way to  the
post-Newtonian limit. As matter of fact one obtains a
different gravitational potential with respect to the ordinary
Newtonian one and  the fourth order corrections
in term of the $v/c$ ratio (PPN level) have to be evaluated in a
 different  way. In other words, considering
Brans-Dicke and  O'Hanlon theories, despite of their similar
structure, will imply completely different predictions in Newtonian
limit. Such an achievement represents a significant argument
against the claim that fourth order gravity models can be ruled
out only on the bases of the analogy with Brans-Dicke PPN
parameters.

Another important point has to be considered. The
PPN-parameters $\gamma$ and $\beta$, in the GR context, are intended to parameterize the deviations
from the Newtonian behaviour of the gravitational
potentials. They are defined according to the standard Eddington metric

\begin{eqnarray}
g_{00}=1-\frac{r_g}{|\mathbf{x}|}+\frac{\beta}{2}\frac{{r_g}^2}{|\mathbf{x}|^2}\,,
\end{eqnarray}
\begin{eqnarray}
g_{ij}=-\biggl(1+\gamma\frac{r_g}{|\mathbf{x}|}\biggr)\delta_{ij}\,.
\end{eqnarray}
In particular, the PPN parameter $\gamma$ is related with the
second order correction to the gravitational potential while
$\beta$ is linked with the fourth order  perturbation in
$v/c$. Actually, if we consider the limit $f(R)\rightarrow R$ from
Eqs. (\ref{gsol1}) and (\ref{gsol2}), we have

\begin{eqnarray}
g_{00}=1-\frac{4}{3a}\frac{r_g}{|\mathbf{x}|}\,,
\end{eqnarray}
\begin{eqnarray}
g_{ij}=-\biggl(1+\frac{2}{3a}\frac{r_g}{|\mathbf{x}|}\biggr)\delta_{ij}\,.
\end{eqnarray}
Since $a$ is an arbitrary constant, in order to match the
Newtonian gravitational potential of GR, we should fix $a=4/3$.
This assumption implies

\begin{eqnarray}
g_{00}=1-\frac{r_g}{|\mathbf{x}|}\,,
\end{eqnarray}
\begin{eqnarray}
g_{ij}=-\biggl(1+\frac{1}{2}\frac{r_g}{|\mathbf{x}|}\biggr)\delta_{ij}\,.
\end{eqnarray}
which suggest that the PPN parameter $\gamma$, in this limit,
is 1/2, that is in striking contrast with GR predictions. Such
a result is not surprising. In fact, the GR limit of
the O'Hanlon theory requires $\phi\sim const$
$V(\phi)\rightarrow 0$ but such approximations induce mathematical
inconsistencies in the field equations of $f(R)$-gravity, once these
have been obtained by a general O'Hanlon Lagrangian. Actually,
this is a general feature of the O'Hanlon theory. In fact it can be easily
demonstrated that the field Eqs. (\ref{fets}) and(\ref{fetstr}) do not reduce to
the standard GR ones since we have\,:

\begin{eqnarray}
R_{\mu\nu}=\frac{\mathcal{X}}{a}\Sigma_{\mu\nu}\,,
\end{eqnarray}
\begin{eqnarray}
\frac{\mathcal{X}}{3}T=0
\end{eqnarray}
but $\Sigma_{\mu\nu}$ components read
$\Sigma_{00}=\frac{2}{3}\rho$ and
$\Sigma_{ij}=\frac{1}{3}\rho\delta_{ij}=\frac{1}{2}\Sigma_{00}\delta_{ij}$
instead  of $S_{00}=\frac{1}{2}\rho$ and
$S_{ij}=\frac{1}{2}\rho\delta_{ij}=S_{00}\delta_{ij}$ as usual. Of
course, $S_{\mu\nu}\,=\,T_{\mu\nu}-\frac{1}{2}g_{\mu\nu}T$ with
$T_{\mu\nu}$ the energy-momentum tensor of matter and Einstein
equations are written in the form
\begin{equation}
R_{\mu\nu}\,=\,\mathcal{X}\,S_{\mu\nu}\,.
\end{equation}
Such a pathology is in order even when the GR limit is recovered
from the  Brans-Dicke theory. In such a case, in order to
match the Hilbert-Einstein Lagrangian, one needs $\phi\sim const$
and $\omega=0$, the immediate consequence is that the PPN
parameter $\gamma$ turns out to be $1/2$, while it is well known
that Brans-Dicke model fulfils the low energy limit prescriptions of GR in
the limit $\omega\rightarrow \infty$. Even in this case, the
problem, with respect to the GR prediction, is that the GR limit
of the model introduces inconsistencies in the field equations. In
other words, it is not possible to impose the same transformation
which leads BD theory into GR at the Lagrangian level on the
solutions and the observables obtained by solving field equations
descending from the general Lagrangian. The relevant aspect of this
discussion is that considering a $f(R)$-model, in analogy with the
O'Hanlon theory and then supposing that the self-interaction
potential is negligible, introduces a pathological behaviour on  the solutions leading to the  PPN parameter $\gamma
= 1/2$. This is what happens when an effective approximation
scheme is introduced in the field equations in order to calculate
the weak field limit of fourth order gravity by means of
Brans-Dicke model. Such a result seems, from another point of
view, to enforce the claim that fourth order gravity models have
to be carefully investigated in this limit and their analogy with
scalar-tensor gravity should be  considered accordingly.

\section{The conformal transformations ad the Newtonian limit}\label{4}

Along the paper we have discussed the Newtonian limit of
$f(R)$- gravity  in term of scalar-tensor gravity
rigorously remaining in the Jordan frame.  In this section,  we discuss
 the Newtonian limit  when a conformal
transformation is applied to the  O'Hanlon theory. In other words
we discuss fourth order gravity models in the Einstein frame in
place of the Jordan one when a redefinition of the metric in the
conformal sense is performed.  A
scalar-tensor gravity theory is, in some sense, a generalization of both  the Brans-Dicke and the O'Hanlon theories, that is

\begin{equation}\label{scaten}
\mathcal{A}_{JF}^{ST}\,=\,\int d^4x
\sqrt{-g}\biggl[F(\phi)R+\omega(\phi)\phi_{;\alpha}\phi^{;\alpha}-V(\phi)+\mathcal{X}
\mathcal{L}_{m}\biggr]\,.
\end{equation}
Such a theory can be transformed by means of a conformal transformation
$\tilde{g}_{\mu\nu}\,=\,A(\mathbf{x})g_{\mu\nu}$, with
$A(\textbf{x})>0$ satisfying the condition
$F(\phi)A(\textbf{x})^{-1}=\Lambda\,\,\epsilon\,\,\mathbb{R}$, as

\begin{equation}\label{confscaten}
\mathcal{A}_{EF}^{ST}\,=\,\int d^4x \sqrt{-\tilde{g}}
\biggl[\Lambda \tilde{R}
+\Omega(\varphi)\varphi_{;\alpha}\varphi^{;\alpha}-W(\varphi)+\mathcal{X}
\tilde{\mathcal{L}}_{m}\biggr]\,.
\end{equation}
The relations between the quantities in the two frames are
\begin{eqnarray}
\left\{\begin{array}{ll}\Omega(\varphi){d\varphi}^2\,=\,\Lambda\biggl[\frac{\omega(\phi)}{F(\phi)}
-\frac{3}{2}\biggl(\frac{d\ln
F(\phi)}{d\phi}\biggr)^2\biggr]{d\phi}^2\\\\W(\varphi)=\frac{\Lambda^2}
{F(\phi(\varphi))^2}V(\phi(\varphi))\\\\\tilde{\mathcal{L}}_m=\frac{\Lambda^2}{F(\phi(\varphi))^2}
\mathcal{L}_m\biggl(\frac{\Lambda\,\tilde{g}_{\rho\sigma}}{F(\phi(\varphi
))}\biggr)\end{array}\right.
\end{eqnarray}
In the case of the
O'Hanlon Lagrangian, (\ref{ohanlon}), i.e.
$F(\phi)\,=\,\phi,\,\,\omega{(\phi)}\,=\,0$, the previous
Lagrangian turns out to be simplified and the transformation rule
between the two scalar fields reads

\begin{equation}\label{phivarphigensol}
\Omega(\varphi){d\varphi}^2\,=\,-\frac{3\Lambda}{2}\frac{{d\phi}^2}{\phi^2}\,\,\,\,\,\,\,\,
\Longrightarrow\,\,\,\,\,\,\,\,
\phi\,=\,k\,e^{\pm\int\sqrt{-\frac{2\Omega(\varphi)}{3
\Lambda}}d\varphi}\,,
\end{equation}
where $k$ is a integration constant. If we suppose
$\Omega(\varphi)\,=\,-\Omega_{0}<0$, we have

\begin{equation}\label{phivarphisol}
\phi\,=\,k\,e^{\pm Y\varphi}\,,
\end{equation}
where $Y=\sqrt{\frac{2\Omega_0}{3\Lambda}}$. The
transformed  action (\ref{ohanlon}) in the Einstein frame is

\begin{eqnarray}\label{ohanlon-EF}
\mathcal{A}_{EF}^{OH}=\int
d^4x\sqrt{-\tilde{g}}\biggl[\Lambda\tilde{R}-\Omega_0\varphi_
{;\alpha}\varphi^{;\alpha}-\frac{\Lambda^2}{k^2}e^{\mp
2Y\varphi}V(k\,e^{\pm
Y\varphi})+\frac{\mathcal{X}\Lambda^2}{k^2}e^{\mp
2Y\varphi}\mathcal{L}_m\biggl(\frac{\Lambda}{k}e^{\mp
Y\varphi}\tilde{g}_{\rho\sigma}\biggr)\biggr]\,.
\end{eqnarray}
The field equations  are now

\begin{eqnarray}
\left\{\begin{array}{ll}
\Lambda\tilde{G}_{\mu\nu}+\frac{1}{2}\frac{\Lambda^2}{k^2}e^{\mp
2Y\varphi}V(k\,e^{\pm
Y\varphi})\tilde{g}_{\mu\nu}-\Omega_0\biggl(\varphi_{;\mu}\varphi_{;\nu}-\frac{1}{2}
\varphi_{;\alpha}\varphi^{;\alpha}\tilde{g}_{\mu\nu}\biggr)=\mathcal{X}\tilde{T}^\varphi_{\mu\nu}
\\\\
2\Omega_0\tilde{\Box}\varphi-\frac{\Lambda^2}{k^2}e^{\mp
2Y\varphi}[\frac{\delta V}{\delta\phi}(k\,e^{\pm Y\varphi})\mp 2 Y
V(k\,e^{\pm
Y\varphi})]+\mathcal{X}\frac{\partial\tilde{\mathcal{L}_{m}}}{\partial\varphi}=0
\\\\
\tilde{R}=-\frac{\mathcal{X}}{\Lambda}\tilde{T}^\varphi+\frac{\Omega_0}{\Lambda}
\varphi_{;\alpha}\varphi^{;\alpha}+\frac{2 \Lambda}{k^2}e^{\mp
2Y\varphi}V(ke^{\pm Y\varphi})
\end{array} \right.
\end{eqnarray}
where the matter tensor, which now coupled with the scalar
field $\varphi$, in the Einstein frame \cite{cqg} reads

\begin{eqnarray}
\tilde{T}^\varphi_{\mu\nu}\,=\,\frac{-2}{\sqrt{-\tilde{g}}}\frac{\delta(\sqrt{-\tilde{g}}\tilde{\mathcal{L}}
_m)}{\delta\tilde{g}^{\mu\nu}}=\frac{\Lambda^2}{k^2}e^{\mp
2Y\varphi}\biggl[\mathcal{L}_m\biggl(\frac{\Lambda}{k}e^{\mp Y
\varphi}\tilde{g}_{\rho\sigma}\biggr)\tilde{g}_{\mu\nu}-2\frac{\delta}{\delta\tilde{g}^{\mu\nu}}\mathcal{L}_m
\biggl(\frac{\Lambda}{k}e^{\mp
Y \varphi}\tilde{g}_{\rho\sigma}\biggr)\biggr]\,,
\end{eqnarray}
and
\begin{eqnarray}
\frac{\partial\tilde{\mathcal{L}_{m}}}{\partial\varphi}=\mp\frac{\Lambda^2Y}{k^2}e^{\mp
2Y\varphi}\biggl[2\mathcal{L}_m\biggl(\frac{ \Lambda}{k}e^{\mp
Y\varphi}\tilde{g}_{\rho\sigma}\biggr)+\frac{\Lambda}{k}e^{\mp
Y\varphi}\tilde{g}_{\rho\sigma}\frac{\delta\mathcal{L}_m}{\delta
g_{\rho\sigma}}\biggl(\frac{\Lambda}{k}e^{\mp Y
\varphi}\tilde{g}_{\rho\sigma}\biggr)\biggr]\,.
\end{eqnarray}
Actually, in order to calculate the Newtonian limit of the model
in the Einstein frame, we can develop the two scalar fields at the
second order $\phi\sim \phi^{(0)}+\phi^{(2)}$ and $\varphi\sim
\varphi^{(0)}+\varphi^{(2)}$ with respect to a background value.
This choice gives the relation\,:

\begin{eqnarray}
\left\{\begin{array}{ll}\varphi^{(0)}\,=\,\pm Y^{-1}\ln{\frac{\phi^{(0)}}{k}}\\\\
\varphi^{(2)}\,=\,\pm
Y^{-1}\frac{\phi^{(2)}}{\phi^{(0)}}\end{array}\right.
\end{eqnarray}
Let us consider the conformal transformation
$\tilde{g}_{\mu\nu}\,=\,\frac{\phi}{\Lambda}g_{\mu\nu}$. From this
relation and considering the (\ref{phivarphisol}) one obtains, if
$\phi^{(0)}=\Lambda$, that

\begin{eqnarray}\label{confg}
\left\{\begin{array}{ll}\tilde{g}^{(2)}_{00}\,=\,g^{(2)}_{00}+\frac{\phi^{(2)}}{\phi^{(0)}}\\\\
\tilde{g}^{(2)}_{ij}\,=\,g^{(2)}_{ij}-\frac{\phi^{(2)}}{\phi^{(0)}}\delta_{ij}\end{array}\right.
\end{eqnarray}
As matter of fact, since $g_{00}^{(2)}\,=\,2 \Phi^{JF}$,
$g_{ij}^{(2)}\,=\,2 \Psi^{JF}\,\delta_{ij}$ and
$\tilde{g}_{00}^{(2)}\,=\,2 \Phi^{EF}$,
$\tilde{g}_{ij}^{(2)}\,=\,2 \Psi^{EF}\,\delta_{ij}$ from
(\ref{confg}) it descends a relevant relation which links the
gravitational potentials of Jordan and Einstein frame\,:

\begin{eqnarray}\label{gravpotconf1}
\left\{\begin{array}{ll}\Phi^{EF}\,=\,\Phi^{JF}+\frac{\phi^{(2)}}{2\phi^{(0)}}\,=
\,\Phi^{JF}\pm\frac{Y}{2}\varphi^{(2)}\\\\
\Psi^{EF}\,=\,\Psi^{JF}-\frac{\phi^{(2)}}{2\phi^{(0)}}\,=\,\Psi^{JF}\mp\frac{Y}{2}
\varphi^{(2)}\end{array}\right.
\end{eqnarray}
If we introduce the variations of two potentials:
$\Delta\Phi\,=\,\Phi^{JF}-\Phi^{EF}$ and
$\Delta\Psi\,=\,\Psi^{JF}-\Psi^{EF}$ we obtain the most relevant
result of this section:

\begin{eqnarray}\label{}
\Delta\Phi\,=\,-\,\Delta\Psi\,=\,-\,\frac{\phi^{(2)}}{2\phi^{(0)}}\,=\,\mp\,\frac{Y}{2}\varphi^{(2)}\,
\propto\,b\,\propto\,f''(R=0)\,.
\end{eqnarray}
From the above expressions, one can notice that there is an evident
difference between the behavior of the two gravitational
potentials in the two frames. Such achievement suggests that, at the
Newtonian level, it is possible to discriminate between the two
 frames.  Specifically, once the gravitational potential is
calculated in the Jordan frame and the dynamical evolution of
$\phi$ is taken into account at the suitable perturbation level,
these can be substituted in the first of Eqs. (\ref{gravpotconf1})
to obtain its evolution in the Einstein frame. The final step is
that the two potentials have to be matched with experimental data in
order to investigate what  the physical solution is.

\section{Conclusions}\label{5}

In this paper, we have  used the analogy between the
$f(R)$-gravity and the O'Hanlon theory to discuss, in the Jordan
frame, the Newtonian limit of the theory.   The main result is that it is not possible to consider the   analogy between $f(R)$-gravity and the Brans-Dicke theory to achieve the correct PPN limit, as done several times in literature,  since the result $\omega_{BD}=0$ implying $\gamma\,=\,1/2$ is a pathology of the theory (both $f(R)$ and Brans-Dicke).  This means that the PPN-parameters have to be redefined accordingly in the Jordan frame of $f(R)$-gravity {\it without}  transforming the theory.

 When we perform the GR limit of $f(R)$-gravity, the correspondence is
needed at  any level (Lagrangian, field
equations and solutions)  between the GR and the $f(R)$-gravity, in particular as soon as $f(R)\sim R^{1+\epsilon}$ with $\epsilon\rightarrow 0$.
This means that the  statement $\gamma\,=\,1/2$ for any $f(R)$ \cite{ppn-bad}
is not correct since the transformation in terms of Brans-Dicke theory does not work .
Supposing  to modify the
Hilbert-Einstein Lagrangian,  the correction to the solutions can
not produce the same displacement from Schwarzschild solution for
any $f(R)$-gravity and  such displacement could not be
independent from the analytical form of $f(R)$. In fact when
$f(R)\rightarrow R$ the solutions of field equations are not the
solutions of GR.
Furthermore the Eddington parameterization \cite{edd}  is based
 on the hypothesis that metric has to match  second order differential field
equations. This  means that gravitational potential  admits the
same Green function of the Newtonian theory \cite{will_et_all}).
In $f(R)$-gravity case, field equations are fourth-order in metric
approach.  The field equations, in Newtonian limit, admit
Yukawa-like corrections  and  the standard Eddington
parameterization cannot work \cite{noi-prd,stelle,bertolami}.
 On the other hand, in order to compare results in Einstein frame and Jordan frame,
 one can perform the Newtonian limit in both frames and then compare the solutions.
 Immediately it emerges that results are different. As final remark,
 it is worth saying that $f(R)$-gravity can fully evade Solar system tests but results have to be carefully
 analyzed in the right frame. Forthcoming experiments could
 clearly give indications in this sense \cite{yunes}.


\begin{thebibliography}{99}
\bibitem{odintsov}
S. Nojiri and S.D. Odintsov, Int. J. Geom. Meth. Mod. Phys.  {\bf
4}, 115 (2007); S. Capozziello, M. Francaviglia, Gen. Rel.
Grav.{\bf 40}, 357 (2008); T. P. Sotiriou,  V. Faraoni arXiv:
0805.1726 [gr-qc] (2008).

\bibitem{ppn-bad}
  T. Chiba, Phys. Lett. B {\bf 575}, 1 (2003);
  M. E. Soussa, R. P. Woodard, Gen. Rel. Grav. {\bf 36}, 855 (2004);
  G.J. Olmo, Phys. Rev. Lett. {\bf 95}, 261102 (2005);
  A.L. Erickcek, T.L. Smith, M. Kamionkowski, Phys. Rev. {\bf D 74}, 121501 (2006);
  T. Chiba, T.L. Smith and A.L. Erickcek, Phys.\ Rev.\  D {\bf 75}, 124014
  (2007).

\bibitem{bransdicke}
  C. Brans and R.H. Dicke, Phys.\ Rev.\  {\bf 124}, 925 (1961).

\bibitem{noi-prd}
  S. Capozziello, A. Stabile, A. Troisi, Phys. Rev. {\bf D 76}, 104019 (2007);
  S. Capozziello, A. Stabile, Class. Quant. Grav. {\bf 26}, 085019 (2009).

\bibitem{altri-ppn-ok}
P.Havas, Gen.Rel. Grav. {\bf 8}, 631 (1977);
  R. Dick, Gen. Rel. Grav. {\bf 36}, 217 (2004);
  I. Navarro, K. Van Acoleyen, Phys. Lett. {\bf B 622}, 1 (2005);
  V. Faraoni, Phys. Rev. {\bf 74}, 023529 (2006);
  T.P. Sotiriou, Gen. Rel. Grav. {\bf 38}, 1407 (2006);
  A. J. Bustelo, D.E. Barraco, Class. Quant. Grav. {\bf 24}, 2333
  (2007).

\bibitem{ohanlon}
  J. O'Hanlon, Phys. Rev. Lett. {\bf 29, 2}, 137 (1972).

\bibitem{weinberg}
  S. Weinberg, \emph{Gravitation and Cosmology}, Wiley 1972, New York.

\bibitem{will}
  C.M  Will, {\it Theory and experiment in gravitational physics},
  Cambridge University Press, Cambridge, 1993 London, 2nd edition; C.M. Will, Living Rev.\ Rel.\  {\bf 4}, 4 (2001).

\bibitem{post-mink}
  S. Capozziello, A. Stabile, A. Troisi, arXiv: 1001.0847v1 (2010).

\bibitem{edd}
  A.S. Eddintgton, \emph{The Mathematical Theory of
  Relativity}, Cambridge University Press 1922, London.

\bibitem{will_et_all}
  K. Nordtvedt J., Astroph. Jour.  {\bf 161}, 1059,
  (1970); C.M. Will, K. Nordtvedt J., Astroph. Jour.  {\bf 177}, 757,
  (1972); K. Nordtvedt J., C.M. Will, Astroph. Jour. {\bf 177}, 775,
  (1972); C.M. Will, Astroph. Jour.  {\bf 185},31, (1973).

\bibitem{stelle}
  K.S. Stelle, Gen. Rev. Grav. {\bf 9}, 343, (1978).

\bibitem{bertolami}
J. Paramos, O. Bertolami Phys. Rev. D {\bf 77}, 084014 (2008).

\bibitem{cqg}
S. Capozziello, R. de Ritis, A.A. Marino,   Class. Quantum Grav.
{\bf 14}  3243, (1997).

\bibitem{yunes}
C. M. Will, N. Yunes,  Class. Quantum Grav. {\bf 21}  4367,
(2004).
\end{thebibliography}
\end{document}